\documentclass[11pt]{article}%
\usepackage{amsfonts}
\usepackage{amsmath}
\usepackage{amssymb}
\usepackage{graphicx}%
\providecommand{\U}[1]{\protect\rule{.1in}{.1in}}
\begin{document}

\title{Short Time Quantum Propagator and Bohmian Trajectories}
\author{Maurice de Gosson\thanks{Financed by the Austrian Research Agency FWF
(Projektnummer P20442-N13).}\\\textit{Universit\"{a}t Wien,}\\\textit{Fakult\"{a}t f\"{u}r Mathematik, NuHAG}\\\textit{Wien 1090 }
\and Basil Hiley\\\textit{University of London}\\\textit{Birkbeck College, Theoretical Physics Unit}\\\textit{London WC1E\ 7HX}}
\maketitle

\begin{abstract}
We begin by giving correct expressions for the short-time action; following
the work of one of us and Makri--Miller. We use these estimates to derive a
correct expression modulo $\Delta t^{2}$ for the quantum propagator and we
show that the quantum potential is negligible modulo $\Delta t^{2}$ for a
point source. We finally prove that this implies that the quantum motion is
classical for very short times.

\end{abstract}


\section{Introduction}

Explicit approximate expressions for the short-time action play an essential
role in various aspects of quantum mechanics (for instance the Feynman path
integral, or semi-classical mechanics), and so does the associated Van Vleck
determinant. Unfortunately, as already observed by Makri and Miller
\cite{makmil1,makmil2}, the literature seems to be dominated by formulas which
are wrong even to the first order of approximation! Using the calculations of
Makri and Miller, which were independently derived by one of us in \cite{ICP}
using a slightly different method, we show that the correct approximations
lead to precise estimates for the short-time Bohmian quantum trajectories for
an initially sharply located particle. We will see that these trajectories are
classical to the second order in time, due to the vanishing of the quantum
potential for small time intervals.

In this paper we sidestep the philosophical and ontological debate around the
\textquotedblleft reality\textquotedblright\ of Bohm's trajectories and rather
focus on the mathematical issues.

\section{Bohmian trajectories}

Consider a time-dependent Hamiltonian function%
\begin{equation}
H(x,p,t)=\sum_{j=1}^{n}\frac{p_{j}^{2}}{2m_{j}}+U(x,t) \label{HAM}%
\end{equation}
and the corresponding quantum operator%
\begin{equation}
\widehat{H}(x,-i\hbar\nabla_{x},t)=\sum_{j=1}^{n}\frac{-\hbar^{2}}{2m_{j}%
}\frac{\partial^{2}}{\partial x_{j}^{2}}+U(x,t). \label{qham}%
\end{equation}
The associated Schr\"{o}dinger equation is%
\begin{equation}
i\hbar\frac{\partial\Psi}{\partial t}=\widehat{H}(x,-i\hbar\nabla_{x}%
,t)\Psi\text{ \ , \ }\Psi(x,0)=\Psi_{0}(x). \label{scheqn1}%
\end{equation}
Let us write $\Psi$ in polar form $Re^{i\Phi/\hbar};$ here $R=R(x,t)\geq0$ and
$\Phi=S(x,t)$ are real functions. On inserting $Re^{iS/\hbar}$ into
Schr\"{o}dinger's equation and separating real and imaginary parts, one sees
that the functions $R$ and $S$ satisfy, at the points $(x,t)$ where
$R(x,t)>0$, the coupled system of non-linear partial differential equations%
\begin{gather}
\frac{\partial S}{\partial t}+\sum_{j=1}^{n}\frac{1}{2m_{j}}\left(
\frac{\partial S}{\partial x_{j}}\right)  ^{2}+U(x,t)-\sum_{j=1}^{n}%
\frac{\hbar^{2}}{2m_{j}R}\frac{\partial^{2}R}{\partial x_{j}^{2}%
}=0\label{bohm1}\\
\frac{\partial R^{2}}{\partial t}+\sum_{j=1}^{n}\frac{1}{m_{j}}\frac{\partial
}{\partial x_{j}}\left(  R^{2}\frac{\partial\Phi}{\partial x_{j}}\right)  =0.
\label{bohm2}%
\end{gather}
The crucial step now consists in recognizing the first equation as a
Hamilton--Jacobi equation, and the second as a continuity equation. In fact,
introducing the quantum potential%
\begin{equation}
Q^{\Psi}=-\sum_{j=1}^{n}\frac{\hbar^{2}}{2m_{j}R}\frac{\partial^{2}R}{\partial
x_{j}^{2}} \label{QP}%
\end{equation}
(Bohm and Hiley \cite{bohi93}) and the velocity field%
\begin{equation}
v^{\Psi}(x,t)=\left(  \frac{1}{m_{1}}\frac{\partial\Phi}{\partial x_{1}%
},...,\frac{1}{m_{n}}\frac{\partial\Phi}{\partial x_{n}}\right)  \label{VELO}%
\end{equation}
the equations (\ref{bohm1}) and (\ref{bohm2}) become
\begin{gather}
\frac{\partial\Phi}{\partial t}+H(x,\nabla_{x}\Phi,t)+Q^{\Psi}%
(x,t)=0\label{bohm3}\\
\frac{\partial\rho}{\partial t}+\operatorname{div}(\rho v^{\Psi})=0\text{ \ ,
\ }\rho=R^{2} \label{bohm4}%
\end{gather}
The main postulate of the Bohmian theory of motion is that particles follow
quantum trajectories, and that these trajectories are the solutions of the
differential equations%
\begin{equation}
\dot{x}_{j}^{\Psi}=\frac{\hbar}{m_{j}}\operatorname{Im}\frac{1}{\Psi}%
\frac{\partial\Psi}{\partial x_{j}}. \label{xdot}%
\end{equation}

The phase space interpretation is that the Bohmian trajectories are determined
by the equations
\begin{equation}
\dot{x}_{j}^{\Psi}=\frac{1}{m_{j}}p_{j}^{\Psi}\text{ \ , }\dot{p}_{j}^{\Psi
}=-\frac{\partial U}{\partial x_{j}}(x^{\Psi},t)-\frac{\partial Q^{\Psi}%
}{\partial x_{j}}(x^{\Psi},t). \label{Bohmeq}%
\end{equation}
It is immediate to check that these are just Hamilton's equations for the
Hamiltonian function
\begin{equation}
H^{\Psi}(x,p,t)=\sum_{j=1}^{n}\frac{p_{j}^{2}}{2m_{j}}+U(x,t)+Q^{\Psi}(x,t)
\label{HAMQ}%
\end{equation}
which can be viewed as a perturbation of the original Hamiltonian $H$ by the
quantum potential $Q^{\Psi}$ (see Holland \cite{Holland1,Holland2} for a
detailed study of quantum trajectories in the context of Hamiltonian mechanics).

The Bohmian equations of motion are \emph{a priori} only defined when $R\neq0$
(that is, outside the nodes of the wavefunction); this will be the case in our
constructions since for sufficiently small times this condition will be
satisfied by continuity if we assume that it is case at the initial time.

An important feature of the quantum trajectories is that they cannot cross;
thus there will be no conjugate points like those that complicate the usual
Hamiltonian dynamics.

\section{The Short-Time Propagator}

The solution $\Psi$ of Schr\"{o}dinger's equation (\ref{scheqn1}) can be
written%
\[
\Psi(x,t)=\int K(x,x_{0};t)\Psi_{0}(x_{0})dx_{0}%
\]
where the kernel $K$ is the \textquotedblleft quantum
propagator\textquotedblright:%
\[
K(x,x_{0};t)=\langle x|\exp(-i\widehat{H}t/\hbar)|x_{0}\rangle.
\]
Schr\"{o}dinger's equation (\ref{scheqn1}) is then equivalent to%
\begin{equation}
i\hbar\frac{\partial K}{\partial t}=\widehat{H}(x,-i\hbar\nabla_{x}%
,t)K\ ,\ K(x,x_{0};0)=\delta(x-x_{0}) \label{K}%
\end{equation}
where $\delta$ is the Dirac distribution. Physically this equation describes
an isotropic source of point-like particle emanating from the point $x_{0}$ at
initial time $t_{0}=0$. We want to find an asymptotic formula for $K$ for
short time intervals $\Delta t$. Referring to the usual literature, such
approximations are given by expressions of the type
\[
K(x,x_{0};\Delta t)=\left(  \frac{1}{2\pi i\hbar}\right)  ^{n/2}\sqrt
{\rho(x,x_{0};\Delta t)}\exp\left(  \frac{i}{\hbar}S(x,x_{0};\Delta t)\right)
\]
where $S(x,x_{0};\Delta t)$ is the action along the classical trajectory from
$x_{0}$ to $x$ in time $\Delta t$ and%
\[
\rho(x,x_{0};\Delta t)=\det\left(  -\frac{\partial^{2}S(x,x_{0};\Delta
t)}{\partial x_{j}\partial x_{k}}\right)  _{1\leq j,k\leq n}%
\]
is the corresponding Van Vleck determinant. It is then common practice
(especially in the Feynman path integral literature) to use the following
\textquotedblleft midpoint approximation\textquotedblright\ for the generating
function $S$: \
\begin{equation}
S(x,x_{0};\Delta t)\approx\sum_{j=1}^{n}\frac{m_{j}}{2\Delta t}(x_{j}%
-x_{0})^{2}-\frac{1}{2}(U(x,t_{0})+U(x_{0},t_{0}))\Delta t \label{r1}%
\end{equation}
or, worse,
\begin{equation}
S(x,x_{0};t,t_{0})\approx\sum_{j=1}^{n}\frac{m_{j}}{2\Delta t}(x_{j}%
-x_{0})^{2}-(U(\tfrac{1}{2}(x+x_{0}),t_{0})\Delta t \label{r2}%
\end{equation}
However, as already pointed out by Makri and Miller \cite{makmil1,makmil2},
these \textquotedblleft approximations\textquotedblright\ are wrong; they fail
to be correct even to first order in $\Delta t$! In fact, Makri and Miller,
and one of us \cite{ICP} have shown independently, and using different
methods, that the correct asymptotic expression for the generating function is
given by%
\begin{equation}
S(x,x_{0};\Delta t)=\sum_{j=1}^{n}\frac{m_{j}}{2\Delta t}(x_{j}-x_{0}%
)^{2}-\widetilde{U}(x,x_{0})\Delta t+O(\Delta t^{2}) \label{correct}%
\end{equation}
where $\widetilde{U}(x,x_{0},0)$ is the average value of the potential over
the straight line joining $x_{0}$ at time $t_{0}$ to $x$ at time $t$ with
constant velocity:%
\begin{equation}
\widetilde{U}(x,x_{0})=\int_{0}^{1}U(\lambda x+(1-\lambda)x_{0},0)d\lambda.
\label{vtilde}%
\end{equation}

For instance when
\[
H(x,p)=\frac{1}{2m}(p^{2}+m^{2}\omega^{2}x^{2})
\]
is the one-dimensional harmonic oscillator formula (\ref{correct}) yields the
correct expansion%
\begin{equation}
S(x,x_{0};t)=\frac{m}{2\Delta t}(x-x_{0})^{2}-\frac{m\omega^{2}}{6}%
(x^{2}+xx_{0}+x_{0})\Delta t+O(\Delta t^{2}); \label{wharmo1}%
\end{equation}
the latter can of course be deduced directly from the exact value%
\begin{equation}
S(x,x_{0};t,t_{0})=\frac{m\omega}{2\sin\omega\Delta t}((x^{2}+x_{0})\cos
\omega\Delta t-2xx_{0}) \label{wharmo2}%
\end{equation}
by expanding $\sin\omega\Delta t$ and $\cos\omega\Delta t$ for $\Delta
t\rightarrow0$. This correct expression is of course totally different from
the erroneous approximations obtained by using the \textquotedblleft
rules\textquotedblright\ (\ref{r1}) or (\ref{r2}).

Introducing the following notation,%
\begin{equation}
\widetilde{S}(x,x_{0};\Delta t)=\sum_{j=1}^{n}m_{j}\frac{(x_{j}-x_{0})^{2}%
}{2\Delta t}-\widetilde{U}(x,x_{0})\Delta t , \label{wtilda}%
\end{equation}
leads us to the Makri and Miller approximation (formula (17c) in
\cite{makmil1}) for the short-time propagator:%
\begin{equation}
K(x,x_{0};\Delta t)=\left(  \frac{1}{2\pi i\hbar}\right)  ^{n/2}\sqrt
{\rho(x,x_{0};\Delta t)}\exp\left(  \frac{i}{\hbar}\widetilde{S}%
(x,x_{0};\Delta t)\right)  +O(\Delta t^{2}) \label{makri1}%
\end{equation}
where
\[
\rho(x,x_{0};\Delta t)=\det\left(  -\frac{\partial^{2}\widetilde{S}%
(x,x_{0};\Delta t)}{\partial x_{j}\partial x_{k}}\right)  _{1\leq j,k\leq n}.
\]
It turns out that this formula can be somewhat improved. The Van Vleck
determinant $\rho(x,x_{0};\Delta t)$ is explicitly given, taking formula
(\ref{wtilda}) into account, by
\[
\rho(x,x_{0};\Delta t)=\det\left(  -\frac{1}{\Delta t}M-\widetilde{U}%
_{x,x_{0}}^{\prime\prime}(x,x_{0})\Delta t\right)
\]
where $M$ is the mass matrix (the diagonal matrix with positive entries the
masses $m_{j}$) and
\[
\widetilde{U}_{x,x_{0}}^{\prime\prime}=\left(  -\frac{\partial^{2}%
\widetilde{U}(x,x_{0})}{\partial x_{j}\partial x_{k}}\right)  _{1\leq j,k\leq
n}.
\]
Writing%
\begin{align*}
\left(  -\frac{1}{\Delta t}M-\widetilde{U}_{x,x_{0}}^{\prime\prime}%
(x,x_{0})\Delta t\right)   &  =-\frac{1}{\Delta t}M[I_{n\times n}%
-M^{-1}\widetilde{U}_{x,x_{0}}^{\prime\prime}(x,x_{0})\Delta t^{2}]\\
&  =-\frac{1}{\Delta t}M[I_{n\times n}+O(\Delta t^{2})],
\end{align*}
we have by taking the determinant of both sides%
\[
\rho(x,x_{0};\Delta t)=\frac{m_{1}\cdot\cdot\cdot m_{n}}{(\Delta t)^{n}}%
\det\left(  I_{n\times n}+O(\Delta t^{2})\right)  .
\]
Noting that $\det\left(  I_{n\times n}+O(\Delta t^{2})\right)  =1+O(\Delta
t^{2})$, we thus have%
\begin{equation}
\rho(x,x_{0};\Delta t)=\frac{m_{1}\cdot\cdot\cdot m_{n}}{(\Delta t)^{n}%
}(1+O(\Delta t^{2})). \label{rho2}%
\end{equation}
Writing%
\begin{equation}
\widetilde{\rho}(\Delta t)=\frac{m_{1}\cdot\cdot\cdot m_{n}}{(\Delta t)^{n}}
\label{rhofree}%
\end{equation}
which is just the Van Vleck density for the free particle Hamiltonian. We thus
have%
\begin{equation}
\rho(x,x_{0};\Delta t)=\widetilde{\rho}(\Delta t)+O(\Delta t^{2})
\label{rhorho}%
\end{equation}
and hence we can rewrite formula (\ref{makri1}) as
\begin{equation}
K(x,x_{0};\Delta t)=\left(  \frac{1}{2\pi i\hbar}\right)  ^{n/2}%
\sqrt{\widetilde{\rho}(\Delta t)}\exp\left(  \frac{i}{\hbar}\widetilde{S}%
(x,x_{0};\Delta t)\right)  +O(\Delta t^{2}). \label{good}%
\end{equation}
We will see below that this formula allows an easy study of the quantum
potential for $K$.

\section{Short-Time Bohmian Trajectories}

Let us determine the quantum potential $Q$ corresponding to the propagator
$K=K(x,x_{0};t)$ using the asymptotic formulas above. Recall that it describes
an isotropic source of point-like particle emanating from the point $x_{0}$ at
initial time $t_{0}=0$. We have, by definition,%
\[
Q=-\sum_{j=1}^{n}\frac{\hbar^{2}}{2m_{j}\sqrt{\rho}}\frac{\partial^{2}%
\sqrt{\rho}}{\partial x_{j}^{2}}%
\]
which we can rewrite%
\[
Q=-\frac{\hbar}{2}^{2}\frac{M^{-1}\nabla_{x}\cdot\nabla_{x}\sqrt{\rho}}%
{\sqrt{\rho}}%
\]
where $M$ is the mass matrix defined above. We have, using (\ref{rhorho}),%
\[
\sqrt{\rho}=\sqrt{\widetilde{\rho}(\Delta t)}(1+O(\Delta t^{2}))
\]
and hence%
\[
\frac{\partial^{2}\sqrt{\rho}}{\partial x_{j}^{2}}=O((\Delta t)^{2})).
\]
From this it follows that the quantum potential associated with the propagator
satisfies%
\begin{equation}
Q(x,x_{0};\Delta t)=O(\Delta t^{2}). \label{qk}%
\end{equation}

The discussion above suggests that the quantum trajectory of a sharply located
particle should be identical with the classical (Hamiltonian) trajectory for
short times. Let us show this is indeed the case. If we want to monitor the
motion of a single, we have of course to specify its initial momentum which
gives its direction of propagation at time $t_{0}=0$; we set%
\begin{equation}
p(0)=p_{0}. \label{pop}%
\end{equation}
In view of formula (\ref{xdot}), the trajectory in position space is obtained
by solving the system of differential equations
\begin{equation}
\dot{x}=\hbar\operatorname{Im}\frac{M^{-1}\nabla_{x}K}{K}\text{ \ ,
\ }x(0)=x_{0}.\text{\ }%
\end{equation}
Replacing $K$ with its approximation%
\[
\widetilde{K}(x,x_{0};\Delta t)=\left(  \frac{1}{2\pi i\hbar}\right)
^{n/2}\sqrt{\widetilde{\rho}(\Delta t)}\exp\left(  \frac{i}{\hbar
}\widetilde{S}(x,x_{0};\Delta t)\right)
\]
we have, since $K-\widetilde{K}=O(\Delta t^{2})$ in view of (\ref{good}),%
\[
\dot{x}=\hbar\operatorname{Im}\frac{M^{-1}\nabla_{x}\widetilde{K}%
}{\widetilde{K}}+O(\Delta t^{2}).
\]
A straightforward calculation, using the expression (\ref{wtilda}) for the
approximate action $\widetilde{S}(x,x_{0};\Delta t)$, leads to (\textit{cf}.
the proof of Lemma 248 in \cite{ICP}) the relation%
\begin{equation}
\dot{x}(\Delta t)=\frac{x(\Delta t)-x_{0}}{\Delta t}-M^{-1}\nabla
_{x}\widetilde{U}(x(\Delta t),x_{0})\Delta t+O(\Delta t^{2}). \label{xdelta}%
\end{equation}
This equation is singular at time $t=0$ hence the initial condition
$x(0)=x_{0}$ is not sufficient for finding a unique solution; this is of
course consistent with the fact that (\ref{xdelta}) describes an arbitrary
particle emanating from $x_{0}$; to single out one quantum trajectory we have
to use the additional condition (\ref{pop}) giving the direction of the
particle at time $t=0$ (see the discussion in Holland \cite{Holland}, \S 6.9).
We thus have%
\[
x(\Delta t)=x_{0}+M^{-1}p_{0}\Delta t+O(\Delta t^{2});
\]
in particular $x(\Delta t)=x_{0}+O(\Delta t)$ and hence, by continuity,%
\[
\nabla_{x}\widetilde{U}(x(\Delta t),x_{0})=\nabla_{x}\widetilde{U}(x_{0}%
,x_{0})+O(\Delta t).
\]
Let us calculate $\nabla_{x}\widetilde{U}(x_{0},x_{0})$. We have, taking
definition (\ref{vtilde}) into account,
\[
\nabla_{x}\widetilde{U}(x,x_{0})=\int_{0}^{1}\lambda\nabla_{x}U(\lambda
x+(1-\lambda)x_{0},0)d\lambda
\]
and hence%
\[
\nabla_{x}\widetilde{U}(x_{0},x_{0})=\int_{0}^{1}\lambda\nabla_{x}%
U(x_{0},0)d\lambda=\frac{1}{2}\nabla_{x}U(x_{0},0).
\]
We can thus rewrite equation (\ref{xdelta}) as
\[
\dot{x}(\Delta t)=\frac{x(\Delta t)-x_{0}}{\Delta t}-\frac{1}{2}M^{-1}%
\nabla_{x}U(x_{0},0)\Delta t+O(\Delta t^{2}).
\]
Let us now differentiate both sides of this equation with respect to $\Delta
t$:%
\begin{equation}
\ddot{x}(t)=\frac{x(t)-x_{0}}{(\Delta t)^{2}}+\frac{\dot{x}(t)}{\Delta
t}-\frac{1}{2}M^{-1}\nabla_{x}U(x_{0},0)+O(\Delta t) \label{x17}%
\end{equation}
that is, replacing $\dot{x}(\Delta t)$ by the value given by (\ref{xdelta}),%
\begin{equation}
\dot{p}(t)=M\ddot{x}(t)=-\nabla_{x}U(x_{0},0)+O(\Delta t). \label{x18}%
\end{equation}
Solving this equation we get
\begin{equation}
p(t)=p_{0}-\nabla_{x}U(x_{0},0)\Delta t+O(\Delta t^{2}). \label{x19}%
\end{equation}
Summarizing, the solutions of the Hamilton equations are given by%
\begin{align}
x(\Delta t)  &  =x_{0}+\frac{p_{0}}{m}\Delta t+O(\Delta t^{2})\label{eq:29}\\
p(\Delta t)  &  =p_{0}-\nabla_{x}U(x_{0},0)\Delta t+O(\Delta t^{2}).
\label{eq:30}%
\end{align}
These equations are, up to the error terms $O(\Delta t^{2})$ the equations of
motion of a classical particle moving under the influence of the potential
$U$; there is no trace of the quantum potential, which is being absorbed by
the terms $O(\Delta t^{2})$. The motion is thus identical with the classical
motion on time scales of order $O(\Delta t^{2})$.

\section{Conclusion}

This result puts Bohm's original perception, which led him to the causal
interpretation, on a firm mathematical footing. He writes \cite{bhdp87}

\begin{quote}
Indeed it had long been known that when one makes a certain approximation
(WKB) Schr\"{o}dinger's equation becomes equivalent to the classical
Hamilton--Jacobi theory. At a certain point I asked myself: What would happen,
in the demonstration of this equivalence, if we did not make this
approximation? I saw immediately that there would be an additional potential,
representing a kind of force, that would be acting on the particle.
\end{quote}

The source of this \textquotedblleft force\textquotedblright\ was the quantum
potential. In our approach we see that while any classical potential acts
immediately, the quantum potential potential does not. From this fact two
consequences of our follow.

Firstly, it gives a rigorous treatment of the \textquotedblleft watched
pot\textquotedblright\ effect. If we keep observing a particle that, if
unwatched, would make a transition from one quantum state to another, will no
longer make that transition. The unwatched transition occurs when the quantum
potential grows to produce the transition. Continuously observing the particle
does not allow the quantum potential to develop so the transition does not
take place. For details see section 6.9 in Bohm and Hiley \cite{bohi93}.

Secondly, in the situation when the quantum potential decreases continuously
with time, the quantum trajectory continuously deforms into a classical
trajectory \cite{bham95}. This means that there is no need to appeal to
decoherence to reach the classical domain.

\end{document}